\documentclass[a4paper,cite,11pt,psnfss]{article}

\usepackage[latin1]{inputenc}
\usepackage{comment}
\usepackage{ulem}
\usepackage{adjustbox}
\usepackage{appendix}
\usepackage{framed,epsfig,pgfplots,amsmath}
\usepackage{tikz}
\usepackage{tikz-feynhand}
\usetikzlibrary{snakes,arrows,shapes,positioning,automata,backgrounds,calc,er,patterns}
\usepackage{caption}
\usepackage{subcaption}
\usepackage{cite}
\usepackage[colorlinks,citecolor=blue]{hyperref}
\usepackage{mathtools, amsmath, amsthm, amssymb,amsfonts}
\usepackage{makecell}
\usepackage{bm}
\usepackage{url}
\usepackage{hhline}
\usepackage{multirow}
\renewcommand{\arraystretch}{1.5}

\usepackage[usenames,dvipsnames,tree]{pstricks}

\definecolor{boxcolor}{RGB}{235,245,255}
\newcommand{\mybox}[1]{\begin{center}\fcolorbox{black}{boxcolor}{\parbox[c]{16.5cm}{#1}}\end{center}}
\newcommand{\inbox}[2]{\begin{tabular}{rl}{\tt In[#1]:= } #2\end{tabular}}
\newcommand{\outbox}[2]{\begin{tabular}{rl}{Prints $\Rightarrow$\hspace{0.85mm}} #2\end{tabular}}
\newcommand{\boxsplit}{\rule{16.5cm}{0.5pt}}
\newcommand{\gibspace}[2]{\begin{tabular}{rl}{\tt \hphantom{aaaaaaa} } #2\end{tabular}}

\usepackage{graphicx,color,pst-plot}

\usepackage{latexsym,amsmath,amsfonts,amssymb,amstext,mathrsfs,upgreek}

\allowdisplaybreaks[1]

\begin{document}

\begin{titlepage}
\begin{flushright}
PSI-PR-23-31, ZU-TH 44/23
\end{flushright}
\begin{flushright}
\end{flushright}

\vfill

\begin{center}

{\Large\bf Multiple Mellin-Barnes integrals and}
\medskip
%{\Large\bf  and} 

{\Large\bf triangulations of point configurations}

 %\vspace{0.5cm}
%{\Large\bf }

\vfill

{\bf Sumit Banik$^{a,b\dagger}$ and Samuel Friot$^{c,d\ddagger}$}\\[1cm]
{$^a$ Physik-Institut, Universitat Zurich,\\ Winterthurerstrasse 190, CH 8057 Zurich, Switzerland}\\[0.5cm]
{$^b$ Paul Scherrer Institut, CH 5232 Villigen PSI, Switzerland}\\[0.5cm]
{$^c$ Universit\'e Paris-Saclay, CNRS/IN2P3, IJCLab, 91405 Orsay, France } \\[0.5cm]
{$^d$ Univ Lyon, Univ Claude Bernard Lyon 1, CNRS/IN2P3, \\
 IP2I Lyon, UMR 5822, F-69622, Villeurbanne, France}\\[0.5cm]
\end{center}
\vfill

\begin{abstract}
We present a novel %systematic and efficient 
technique for the analytic evaluation of multifold Mellin-Barnes (MB) integrals, which commonly appear in physics, as for instance in the calculations of multi-loop multi-scale Feynman integrals. Our approach is based on triangulating a set of points which can be assigned to a given MB integral, and yields the final analytic results in terms of linear combinations of multiple series, each triangulation allowing the derivation of one of these combinations. When this technique is applied to the computation of Feynman integrals, the involved series are of the (multivariable) hypergeometric type. We implement our method in the  \textit{Mathematica} package \texttt{MBConicHulls.wl}, an already existing software dedicated to the analytic evaluation of multiple MB integrals, based on a recently developed computational approach using intersections of conic hulls. The triangulation method is remarkably faster than the conic hulls approach and can thus be used for the calculation of higher-fold MB integrals as we show here by computing triangulations for highly complicated objects such as the off-shell massless scalar one-loop $15$-point Feynman integral whose MB representation has 104 folds. As other applications we show how this technique can provide new results for the off-shell massless conformal hexagon and double box Feynman integrals, as well as for the hard diagram of the two loop hexagon Wilson loop.

\end{abstract}

\vspace{1cm}

\small{$\dagger$ sumit.banik@psi.ch}

\small{$\ddagger$ samuel.friot@universite-paris-saclay.fr}

\end{titlepage}

\section{Introduction}
Mellin-Barnes (MB) integrals are a special type of integrals whose integrand consists of a ratio of products of Euler gamma functions (the latter's arguments being linear combinations of the integration variables as well as, possibly, constant terms) and parameters raised to the power of the integration variables. The integration contours of such integrals follow certain paths in the complex planes of the integration variables which avoid the poles of the integrand. MB integrals find many applications in different branches of physics and mathematics. 
In particle physics, for instance, MB integrals often appear in the evaluation of Feynman integrals \cite{Smirnov:2012gma,Dubovyk:2022obc}. In this context, the original Feynman integral is first converted into an MB integral using standard procedures \cite{Smirnov:2012gma,Dubovyk:2022obc,Gluza:2007rt,Ambre,Belitsky:2022gba} and thereby the focus is only on the MB integral. Some of the subsequent applications include resolving $\epsilon$ singularities of dimensionally regularized Feynman integrals \cite{Smirnov:1999gc,Tausk:1999vh,Czakon:2005rk,Smirnov:2009up}, finding their analytic expressions in terms of hypergeometric functions \cite{Kalmykov:2020cqz},  multiple polylogarithms \cite{Smirnov:2012gma,Vollinga:2004sn}, performing numerical integration \cite{Dubovyk:2019krd}, counting master integrals (in some cases) \cite{Kalmykov:2016lxx}, deriving partial differential equations without relying on integration-by-parts identities \cite{Kalmykov:2012rr}, etc.

MB integrals are also very useful in the theory of multivariable hypergeometric functions. Indeed, one of the primary reasons to study MB integrals, during their early stage, was due to their applications to the theory of hypergeometric functions, where they were recognized as a powerful tool for the derivation of their linear transformations (see for instance the pioneering works \cite{Barnes, W&W} for the simplest case of the Gauss ${}_2F_1$ function, and \cite{KdF,Erdelyi} for the next-to-simplest cases of the generalized ${}_pF_q$ and Appell $F_1, ..., F_4$ functions).

Although MB integrals are widely used, there was no efficient and systematic computational technique for their analytic calculation in the arbitrary multifold case until the recent works \cite{Ananthanarayan:2020fhl,Banik:2022bmk}. In \cite{Ananthanarayan:2020fhl}, for a given $N$-fold MB integral (with fixed $N$), a geometrical approach, mixing specific intersections of conic hulls associated with the MB integrand and results of multidimensional complex analysis, was presented as a first solution to this problem. It was automated in the form of a \textit{Mathematica} package called \texttt{MBConicHulls.wl} \cite{Ananthanarayan:2020fhl}. Direct applications of this new method to the analytic computation of the nine-fold MB representations of highly non-trivial Feynman integrals showed its efficiency: see \cite{Ananthanarayan:2020ncn} for the first computation of the off-shell massless hexagon and double-box conformal Feynman integrals. 
 The conic hulls method and the package were then improved, in order to handle the straight contours cases, in \cite{Banik:2022bmk}. 
However, although very useful, the \texttt{MBConicHulls.wl} package becomes limited in speed when tackling the computation of complicated objects such as those considered in \cite{Ananthanarayan:2020ncn}.
In the present paper, we develop a novel geometrical approach for the analytic evaluation of multifold MB integrals, based on triangulations of configurations of points which, in addition to the potentially new insights in the theory of MB integrals that it can offer, is computationally much more efficient than the conic hull approach. We have checked the latter point with a new version of the \texttt{MBConicHulls.wl} \cite{MBConicHullsGit} package where we have implemented the triangulation procedure by adding a new module incorporating the \texttt{TOPCOM} software \cite{Rambau:TOPCOM:2002}.
As an example of application, we are now able to find all possible series representations of the double-box and hexagon conformal Feynman integrals in a very short time, whereas it would have taken more than a lifetime with the former version of the package. This allowed us to discover that simpler series representations than those previously published in \cite{Ananthanarayan:2020ncn} can be obtained. As another example, we compute the hard diagram of the two loop six-edged Wilson loop in general kinematics \cite{DelDuca:2010zg} and we show that 1471926 different series representations can be obtained from its MB representation. The triangulation approach also makes possible the computation of much more complicated objects than the above mentioned integrals, as we have checked by testing the code on higher-fold MB integrals. As an example, we show in this paper that the computation of triangulations associated with the scalar off-shell massless one-loop $N$-point Feynman integral, for $N$ going up to 15 (for which the corresponding MB representation has 104 folds), is possible. %In fact, we find series solutions of length 25 both for the hexagon and double box. This is simpler than the previous solution which was of length 26 and 44 respectively for hexagon and double box.

The outline of this paper is as follows. In Section \ref{method}, we present the triangulation technique applied to the calculation of $N$-fold MB integrals {and briefly discuss how to use multivariate residues in order to obtain the solutions in terms of multivariable series representations}. In Section \ref{example}, we give the simple example of a two-fold MB integral, in order to illustrate how the technique presented in Section \ref{method} works. In Section \ref{Applications}, we revisit some previously computed Feynman integrals and compare the time-efficiency of the triangulation approach versus the conic hulls computational technique. We present, for the most complicated integrals, namely the hard diagram of the two loop six-edged Wilson loop and the conformal hexagon and double box, new series solutions (which for the latter two are simpler than the ones previously presented in \cite{Ananthanarayan:2020ncn}). We also present in this section the computation of higher-fold MB integrals, by considering the $N$-point integral previously mentioned. Finally, we provide our concluding remarks and discussions in Section \ref{conclude}, and the computer implementation of the triangulation method is detailed in the appendix that follows.

\section{The triangulation method  \label{method}}

A typical $N$-fold MB integral is of the form
\begin{equation} 
    I(x_1,\cdots ,x_N)= \int\limits_{-i \infty}^{+i \infty} \frac{ d z_1}{2 \pi i} \cdots \int\limits_{-i \infty}^{+i \infty}\frac{ d z_N}{2 \pi i}\,\,  \frac{\prod\limits_{i=1}^{k} \Gamma^{a_i}(s_i ({\bf z}))}{\prod\limits_{j=1}^{l} \Gamma^{b_j}(t_j ({\bf z}))}  x^{z_1}_{1} \cdots x^{z_N}_{N}\label{N_MB}
\end{equation}
where ${\bf z}=(z_1,\cdots,z_N)$, $a_i$ and $b_j$ are positive integers, $k\geq N${ (cancellations between numerator and denominator gamma functions are tacitly excluded)}  and the variables $x_1,\cdots ,x_N$ can be complex-valued. The arguments of the gamma functions in the MB integrand are
\begin{align}\label{argument}
    s_i({\bf z}) =\sum\limits_{k=1}^{N}e_{i k}z_k+f_i \, , \hspace{2cm}
    t_j({\bf z}) =\sum\limits_{k=1}^{N}g_{j k}z_k+h_j
\end{align}
where $f_i$ and $h_j$ are real or complex numbers, and the coefficients $e_{ik}$ and $g_{jk}$ are usually integers. In the context of dimensionally regularized Feynman integrals, $f_i$ and $h_j$ are linear combinations of the powers of the propagators and the space-time dimension $D$, $e_{ik}$ and $g_{jk}$ most of the time take the values $1, -1$ or $0$ and $x_1,\cdots ,x_N$ are made of kinematic invariants.

In general, the integration contours in Eq. (\ref{N_MB}) satisfy the following property: they do not split the set of pole of each gamma function of the numerator in different subsets. This is easy to visualize in the 1-fold case where such a rule dictates that the contour separates the left-handed poles of $\Gamma(\cdots+z)$ from the right-handed poles of $\Gamma(\cdots-z)$. In the cases where one would have MB integrals with straight contours which do not separate the left and right-handed poles, one can perform appropriate transformations on the integration variables to separate them, as shown in \cite{Banik:2022bmk}.

For later purpose, we perform a change of the integration variables in order to rewrite Eq. \eqref{N_MB} as
\begin{equation} \label{N_MB_2}
    I(x_1,\cdots ,x_N)= \int\limits_{-i \infty}^{+i \infty} \frac{ d z_1}{2 \pi i} \cdots \int\limits_{-i \infty}^{+i \infty}\frac{ d z_N}{2 \pi i}\,\,  \frac{ \Gamma(-z_1)\cdots\Gamma(-z_N) \prod\limits_{i=N+1}^{k'} \Gamma^{a'_i}(s'_i ({\bf z})) }{\prod\limits_{j=1}^{l} \Gamma^{b'_j}(t'_j ({\bf z}))} x'^{z_1}_{1} \cdots x'^{z_N}_{N}
\end{equation}
where we have pulled out the factors $ \Gamma(-z_1)\cdots\Gamma(-z_N) $ in the numerator\footnote{Note that $k'$ can be different from $k-N$.}. This change of variables always exists for $k\geq N$ and we call Eq.\eqref{N_MB_2} the \textit{canonical form} of the MB representation. In the rest of this article, we assume that MB integrals are written in this form where the gamma function arguments are now
\begin{align}\label{argument_canonical}
    s'_i({\bf z}) =\sum\limits_{k=1}^{N}e'_{i k}z_k+f'_i \, , \hspace{2cm}
    t'_j({\bf z}) =\sum\limits_{k=1}^{N}g'_{j k}z_k+h'_j
\end{align}

For the purpose of the analytic evaluation of  the MB representation in Eq.\eqref{N_MB_2} with our triangulation method, we assign to this integral a set of $k$ points which can be readily extracted from the arguments of the gamma functions of the numerator of its integrand, \textit{i.e.} $ s'_i({\bf z}) $ . This set consists of $N$ points whose {homogeneous} coordinates in the  $k-N$ dimensional Euclidean space are built from the coefficients of the $z_i$ $(i= 1, \cdots, N)$ integration variables of the arguments of numerator's (non-pulled out) gamma functions, \textit{i.e.}
\begin{equation}
P_1 = 
e'_{i 1} \,\, , \hspace{1cm}
P_2 = 
e'_{i 2} \,\, , \hspace{0.5cm} \cdots \hspace{0.5cm}
P_{N} = 
e'_{i N}
\end{equation}
and $k-N$ additional points corresponding to the unit vectors of dimension $k-N$
\begin{equation}
P_{N+1} = \begin{pmatrix}
1 \\ 0 \\ . \\ 0 \\ 0
\end{pmatrix} \, \, , \hspace{1cm}
P_{N+2} = \begin{pmatrix}
0 \\ 1 \\ . \\ 0 \\ 0
\end{pmatrix} \, \, , \hspace{1cm} \cdots \hspace{1cm}
P_{k} = \begin{pmatrix}
0 \\ 0 \\ . \\ 0 \\ 1
\end{pmatrix}
\end{equation}
The above point configuration can be written as a $(k-N) \times k$ matrix where the columns are made of the points $P_i$ $(i=1, \cdots, k)$ and which we denote as the \textit{A}-matrix of the MB integral
\begin{equation}
A=\begin{pmatrix}
P_1 & P_2 & \cdots & P_k
\end{pmatrix}
\end{equation}
In the next step we find all the possible \textit{regular triangulations} of the point configuration $P=\{ P_1 , \cdots, P_k \}$. These triangulations are built from a set of simplices, each simplex being in fact dual to a conic hull in the conic hulls approach. Now, from the remarkable fact that there is a bijective correspondence between the set of possible regular triangulations and the set of relevant intersections of conic hulls in the conic hulls approach, 
 one can associate a set of poles to each triangulation and sum their multivariate residues, as done in \cite{Ananthanarayan:2020fhl}, in order to obtain the series solutions. At the end, each triangulation being used to construct a series representation of the MB integral, one obtains various series solutions which (when the method is applied to the computation of Feynman integrals, or to the derivation of linear transformations of multivariable hypergeometric functions) are in general analytic continuations of each other and converge in different regions of the $(x'_1, \cdots, x'_N)$ $N$-dimensional complex space.

We have implemented the triangulation method in a new version of the \texttt{MBConicHulls.wl} \cite{MBConicHullsGit} \textit{Mathematica} package which can be used for the analytic calculation of $N$-fold MB integrals with an arbitrary (but fixed) number of folds $N\geq 1$,  as described in the appendix of the present paper.

In the next section, we illustrate the above procedure in detail by computing the simple two-fold MB integral associated with the Appell $F_1$ double hypergeometric function.

\section{Appell $F_1$ double hypergeometric function\label{example}} 

The Appell $F_1$ double hypergeometric function provides a nice illustrative example for our triangulation approach. Indeed, as we have already used it in \cite{Ananthanarayan:2020fhl} for the presentation of the conic hulls approach, employing it again here will allow us to easily draw a parallel between these two different techniques.

The MB representation of $F_1$ is two-fold and reads \cite{KdF}:
\begin{align} \label{F1MB}
  F_1(a,b_1,b_2;&c;z_1,z_2)=\frac{\Gamma(c)}{\Gamma(a)\Gamma(b_1)\Gamma(b_2)} \nonumber\\
  &\times\int\limits_{-i \infty}^{+i \infty} \frac{ d s}{2 \pi i}  \int\limits_{-i \infty}^{+i \infty} \frac{ d t}{2 \pi i} (-z_1)^s (-z_2)^t \Gamma(-s)\Gamma(-t)\frac{\Gamma(a+s+t) \Gamma(b_1+s)\Gamma(b_2+t)}{\Gamma(c+s+t)} 
   \end{align}
where the integration contours satisfy the usual requirement: they do not split in subsets the set of poles of each gamma function in the numerator of the integrand.

From this integral it is easy to derive the usual series representation of $F_1$
\begin{align} \label{F1_series}
  F_1(a,b_1,b_2;c;z_1,z_2)=\frac{\Gamma(c)}{\Gamma(a)\Gamma(b_1)\Gamma(b_2)}\sum_{m=0}^{\infty}\sum_{n=0}^{\infty}\frac{\Gamma(a+m+n) \Gamma(b_1+m)\Gamma(b_2+n)}{\Gamma(c+m+n)} \frac{z_1^mz_2^n}{m!n!}
     \end{align}
which converges for $\vert z_1\vert<1$ and $\vert z_2\vert<1$, as well as four analytic continuations of the latter, as shown in \cite{Ananthanarayan:2020fhl} using the conic hull method. Indeed, labelling the five gamma functions of the numerator of the MB integrand by the integers $1,...,5$, it was explicitly shown in \cite{Ananthanarayan:2020fhl} that the following five subsets of conic hulls: 
\begin{equation}
\{C_{1,2}\}, \{C_{1,3},C_{1,5}\}, \{C_{2,3},C_{2,4}\}, \{C_{1,3},C_{3,5},C_{4,5}\} \{C_{2,3},C_{3,4},C_{4,5}\}
\label{Appell_CH}
\end{equation}
 are associated with the five series representations.  The corresponding expressions, which are well-known results, can also be seen as linear transformations of the $F_1$ function and, as described in \cite{Srivastava}, have been derived a long time ago using alternative procedures.

In order to rederive these results from the triangulation approach, we first have to find the configuration of points of the MB representation of $F_1$ given in Eq. \eqref{F1MB}.

Looking at the integrand, one first notes that its numerator is already in the canonical form of Eq. (\ref{N_MB_2}) and, therefore, does not need any change of variables. One can also see that, as there are three ``non-trivial'' gamma functions ($k'=3$) in the numerator of the MB representation of $F_1$ (those involving the parameters $a, b_1$ and $b_2$), the points will belong to the 3-dimensional space. Moreover, as we have two variables of integration in this simple case, there will be only two points to consider in addition to the three others that have to be added automatically.

Therefore, the set of points is 
\begin{equation}
P_{F_1}=\left\{
\begin{pmatrix}
    1 \\
    1 \\
    0
\end{pmatrix},
\begin{pmatrix}
    1 \\
    0 \\
    1
\end{pmatrix},
\begin{pmatrix}
    1 \\
    0 \\
    0
\end{pmatrix},
\begin{pmatrix}
    0 \\
    1 \\
    0
\end{pmatrix},
\begin{pmatrix}
    0 \\
    0 \\
    1
\end{pmatrix}
\right\}
\end{equation}

 \begin{figure}[h!]
\centering
\includegraphics[scale=0.43]{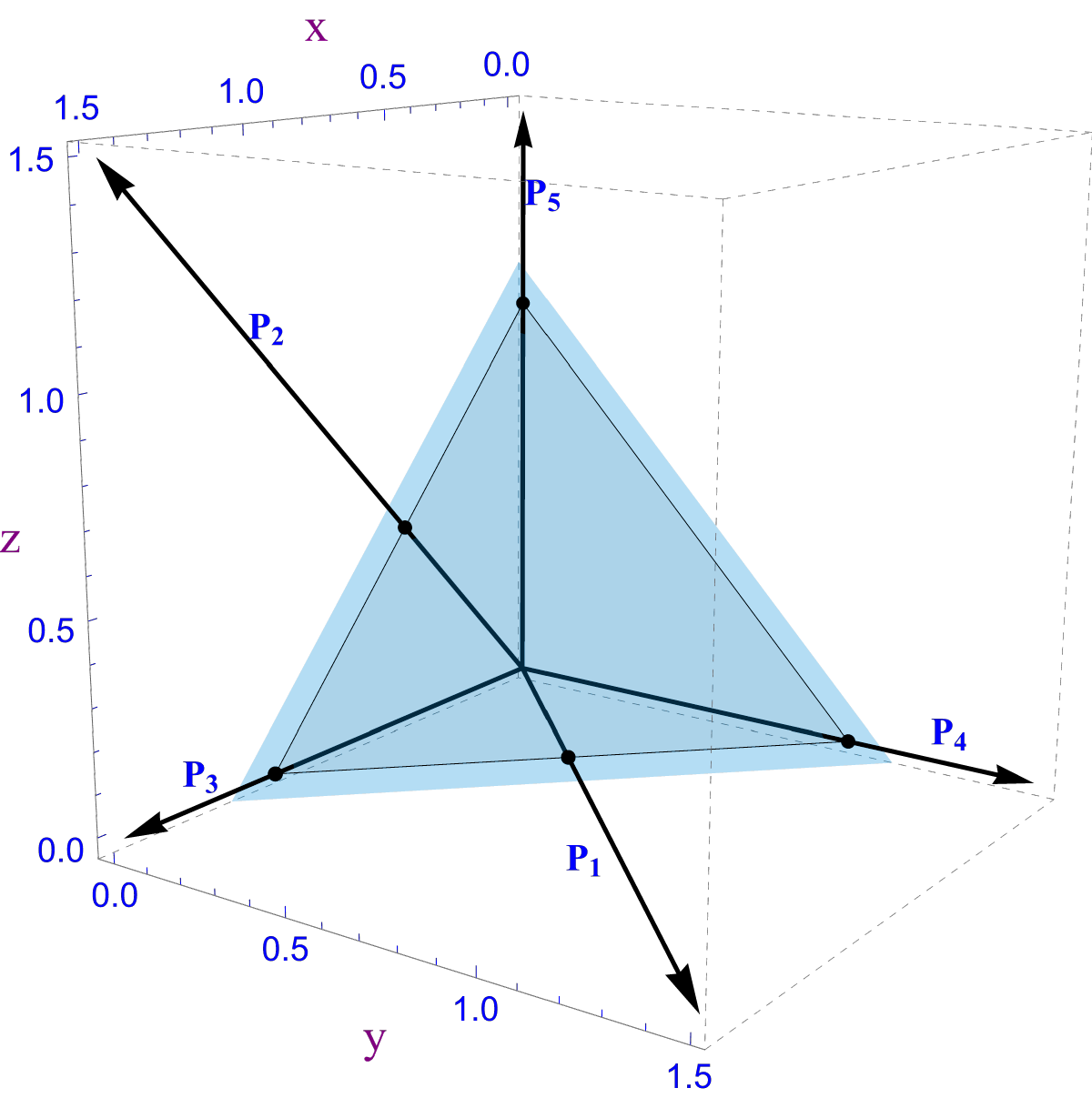}
\caption{Configuration of points $P_{F_1}=\{P_1,...,P_5\}$ associated with the Appell $F_1$ function (homogeneous coordinates).\label{CHF1}}
\end{figure}

 where the first (resp. second) point reflects the $s$ (resp. $t$) dependency of the arguments of the three non-trivial gamma functions of the numerator, whereas the three other points correspond to the automatically added set of points associated with the dimensionality of the configuration space.

\begin{figure}[htb]
     \centering
     \begin{subfigure}[b]{0.3\textwidth}
         \centering
         \includegraphics[scale=0.5]{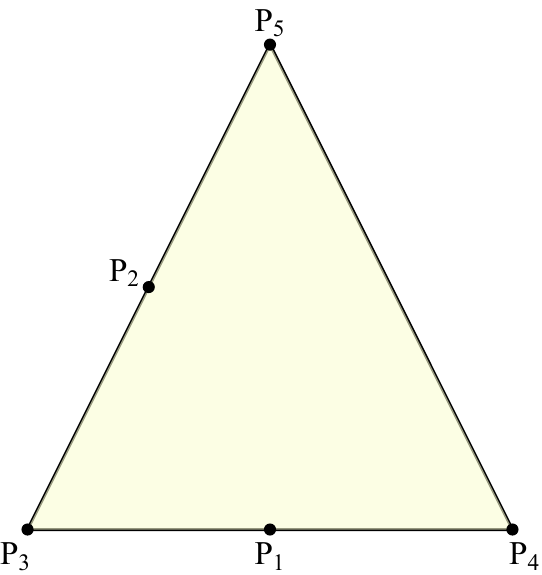}
         \caption{Simplex $\{3,4,5\}$}
     \label{Appell_Tri1}
     \end{subfigure}
     \hfill
     \begin{subfigure}[b]{0.3\textwidth}
         \centering
         \includegraphics[scale=0.5]{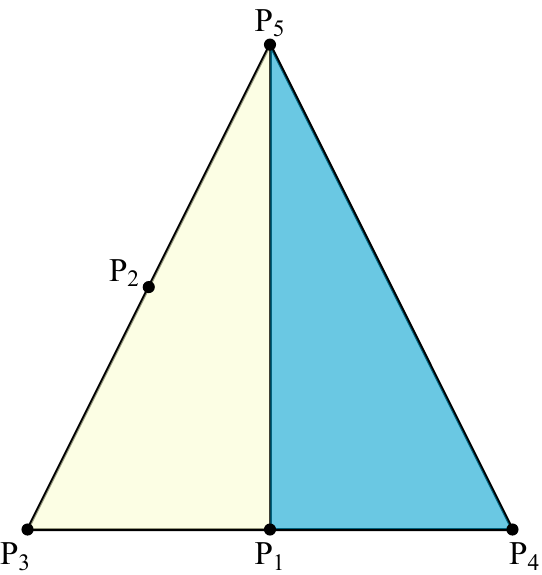}
         \caption{Simplices $\{1,4,5\}, \{1,3,5\}$}
     \label{Appell_Tri2}
     \end{subfigure}
     \hfill
     \begin{subfigure}[b]{0.3\textwidth}
         \centering
         \includegraphics[scale=0.5]{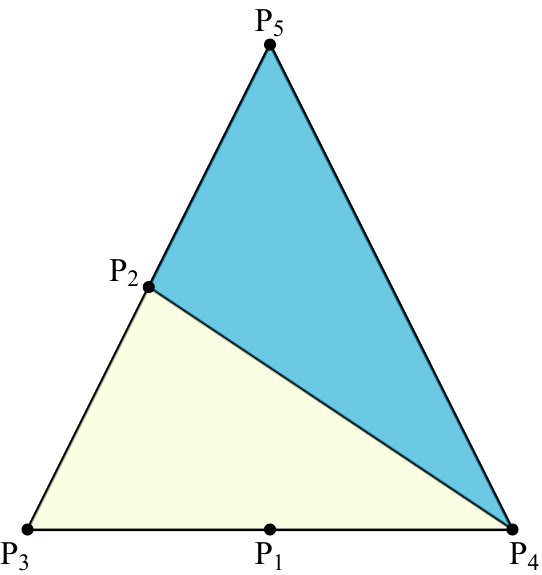}
         \caption{Simplices $\{2,3,4\}, \{2,4,5\}$}
         \label{Appell_Tri3}
     \end{subfigure}
     \hfill
     \vspace{0.5cm}
     \begin{subfigure}[b]{0.45\textwidth}
         \centering
         \includegraphics[scale=0.5]{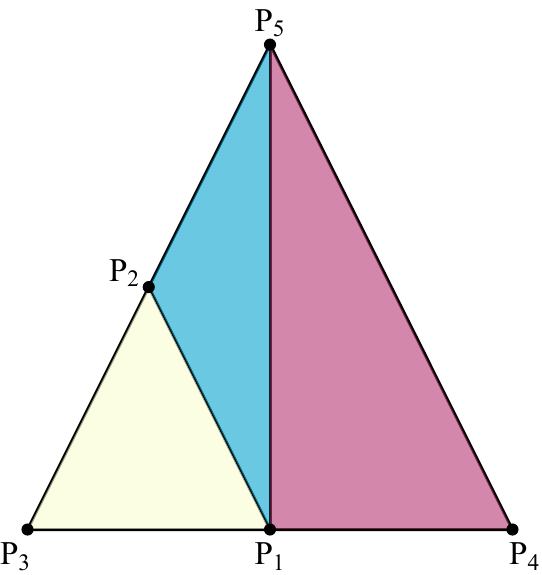}
         \caption{Simplices $\{2,4,5\}, \{1,2,4\}, \{1,2,3\}$}\label{Appell_Tri4}
     \end{subfigure}
     \hspace{-0.5cm}
     \begin{subfigure}[b]{0.45\textwidth}
         \centering
         \includegraphics[scale=0.5]{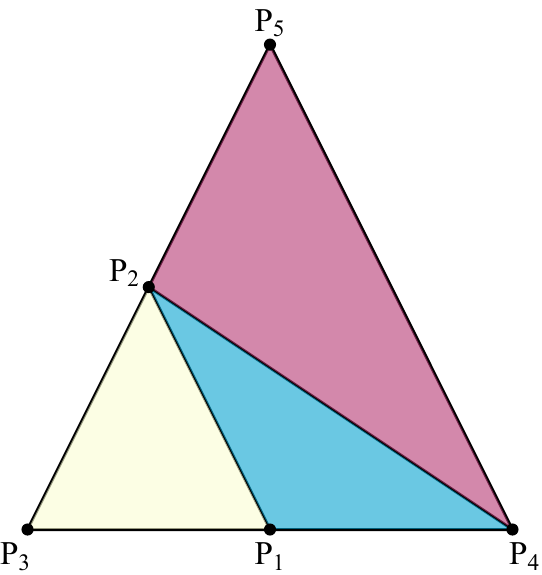}
         \caption{Simplices $\{1,2,4\}, \{1,2,3\}, \{2,4,5\}$}\label{Appell_Tri5}
     \end{subfigure}
        \caption{The five regular triangulations of the configuration of points $P_{F_1}$.\label{Appell_Tri}}
\end{figure}

The set of points $P_{F_1}$ is shown in Fig. \ref{CHF1} where we have associated the labels $P_1,...,P_5$ with these points (we recall that the points have homogeneous coordinates) and the corresponding triangulations are shown in Fig. \ref{Appell_Tri}. Looking at the triangulations,
it is straightforward to deduce, by a direct reading of the notation of the triangulations, the associated subsets of conic hulls shown in \eqref{Appell_CH} as follows. Indeed, in the general case, it is sufficient to find the complements, in the set $\{ 1, \cdots, k'+ N\}$, of the simplices that appear in the triangulations. These complements then give the subsets of conic hulls associated with the series solutions. 

For the particular case of Eq. \eqref{F1MB}, we have $\{ 1, \cdots, k'+ N\}=\{ 1, \cdots, k\}$ with $k=5$ because there are five different numerator gamma functions. Therefore, we find the complements, in $\{1, \cdots, 5 \}$, of each of the simplices of the triangulations in Fig. \ref{Appell_Tri} to get the corresponding subsets of conic hulls. For example, from the triangulation $\mathcal{T}_4= \{ \{2,4,5\} \, , \, \{1,2,4\} \, , \, \{1,2,3\}\}$ in Fig. \ref{Appell_Tri4} we get the complements $ \{ \{1,3\} \, , \, \{ 3,5\} \, , \, \{ 4, 5\} \}$ which are nothing but the fourth subset of conic hulls in (\ref{Appell_CH}). Similarly, one can find the remaining four subsets of conic-hulls from the other triangulations in Fig. \ref{Appell_Tri}. Hence, following \cite{Ananthanarayan:2020fhl} one can subsequently compute the multivariate residues and obtain the various series representations of $F_1$.

\section{Applications to Feynman integrals \label{Applications}}

In this section, we compute higher-fold MB integrals associated with Feynman integrals. We begin with a comparison between the computation times of v.1.1 of \texttt{MBConicHulls.wl}, which is based on conic hulls intersections, and those obtained from the triangulation approach implemented in v.1.2, for several examples having up to nine-fold MB representations. Then we consider MB integrals with a very large number of folds, and test our package by computing triangulations and series representations of the off-shell massless scalar one-loop $N$-point integral, for several values of $N$ going as high as 15 and for which the corresponding MB representations can have up to 104 folds. To our knowledge the corresponding results are new, even for the simplest $N=4$ case. Some details about these results can be found in the ancillary \textit{Mathematica} notebook \texttt{Examples.nb}.

\begin{figure}[htb]
     \centering
     \begin{subfigure}[b]{0.2\textwidth}
         \centering
         \includegraphics[scale=1.4]{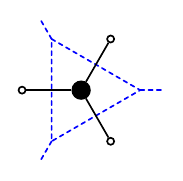}
         \caption{Conformal \\ triangle}
     \label{Conformal_Tri_Diagram}
     \end{subfigure}
     \hspace{0.3cm}
     \begin{subfigure}[b]{0.2\textwidth}
         \centering
         \includegraphics[scale=1.4]{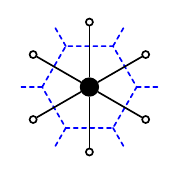}
         \caption{Conformal \\ hexagon}
     \label{Conformal_HX_Diagram}
     \end{subfigure}
      \hspace{0.3cm}
     \begin{subfigure}[b]{0.23\textwidth}
         \centering
         \includegraphics[scale=1.4]{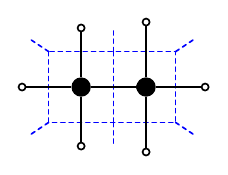}
         \caption{Conformal double-\\ box}
        \label{Conformal_DB_Diagram}
     \end{subfigure}
     \hspace{1cm}
     \begin{subfigure}[b]{0.2\textwidth}
     \hspace{-0.5cm}
         \includegraphics[scale=1.4]{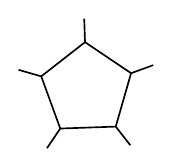}
         \caption{Massless \\ Pentagon}
        \label{Pentagon_Diagram}
     \end{subfigure}
        \caption{One-loop and two-loop Feynman diagrams evaluated using the conic hulls and triangulation methods in Table \ref{Speed}.}
        \label{Feynman_Diagrams}
\end{figure}
\subsection{Comparison of computation times}
We perform the comparison of calculation times on five different Feynman integrals: the off-shell massive conformal triangle which has a three-fold MB representation \cite{Ananthanarayan:2020xpd}, the off-shell massless pentagon in $4-2 \epsilon$ dimensions, whose MB representation has four folds \cite{Banik:2022bmk}, the off-shell massless hexagon and double-box conformal fishnet Feynman integrals in the generic non-resonant $D$ dimensional case and unit resonant four dimensional case \cite{Loebbert:2019vcj} (for interesting recent results on two dimensional fishnet integrals see \cite{Duhr:2022pch})  which both have nine-fold MB representations \cite{Ananthanarayan:2020ncn}, these last two Feynman integrals being related to one another by a differential equation which allows one to check the obtained results, and the hard diagram of the two loop Wilson six-edged Wilson loop \cite{DelDuca:2010zg} (see Fig. \ref{Feynman_Diagrams} for pictures of the first four of these five examples). For the explicit  expressions of their MB representations, we refer the reader to the quoted references. We only give here, as an example, the $A$-matrix of the double-box, which reads 
\begin{align}
A_\text{DB}=                                                     
\left(
\begin{array}{cccccccccccccccc}
 1 & 1 & 1 & 1 & 1 & 1 & 1 & 1 & 1 & 1 & 0 & 0 & 0 & 0 & 0 & 0 \\
 1 & 0 & 0 & 0 & 1 & 0 & 0 & 1 & 1 & 0 & 1 & 0 & 0 & 0 & 0 & 0 \\
 0 & 1 & 0 & 0 & 0 & 1 & 1 & 1 & 0 & 0 & 0 & 1 & 0 & 0 & 0 & 0 \\
 0 & 0 & 1 & 1 & 1 & 1 & 0 & 0 & 0 & 0 & 0 & 0 & 1 & 0 & 0 & 0 \\
 0 & 0 & 0 & -1 & -1 & -1 & -1 & -1 & -1 & 0 & 0 & 0 & 0 & 1 & 0 & 0 \\
 -1 & -1 & -1 & 0 & -1 & -1 & 0 & -1 & 0 & 0 & 0 & 0 & 0 & 0 & 1 & 0 \\
 0 & 0 & 0 & 0 & 0 & 0 & -1 & -1 & -1 & 0 & 0 & 0 & 0 & 0 & 0 & 1 \\
\end{array}
\right)\end{align}
%Similarly, one can find the point configuration of Eq.\eqref{HX} which is a set of $15$ points of dimension $6$.

\begin{table}[h]
  \centering
  \renewcommand{\arraystretch}{1.3}
  \begin{tabular}{|p{2cm}|p{0.7cm}|p{1.4cm}|c|c|c|c|}
    \hline
    {Feynman integral} & {MB folds} & \multirow{2}{1.5cm}{Total solution number}  & \multicolumn{2}{c|}{Conic hulls method} & \multicolumn{2}{c|}{Triangulation method} \\[0.3cm]
    \cline{4-7}
  & &  & One solution & All solutions & One solution & All solutions \\
    \hline
    Conformal triangle & 3 & 14 & 0.186 sec. & 1.44 sec. & 0.543 sec.  & 0.483 sec.  \\[0.3cm] \hline
    Massless pentagon & 5 & 70  & 1.276 sec.  & 1.25 h. & 0.318 sec. & 2.78 sec.  \\[0.3cm] \hline
    Conformal hexagon & 9 & 194160 & 1 min. & -  & 0.489 sec. & 40 min. \\[0.3cm] \hline
    Conformal double-box & 9 & 243186 & 1.9 min. & - & 0.635 sec. & 1.8 h.\\[0.3cm] \hline
    Hard diagram & 8 & 1471926 & 6 min. & - & 1.4 sec. & - \\[0.3cm] \hline
  \end{tabular}
  \caption{Speed comparison of the conic hulls and triangulation methods.}
  \label{Speed}
\end{table}
The results presented in Table \ref{Speed} show the computation times\footnote{On \texttt{Ubuntu 22.04.2} with AMD Ryzen Threadripper Pro 5965WX (24-cores 48-threads) and 128 GB RAM using \textit{Mathematica} \texttt{13.2.1}.} needed for the calculation of one series representation by the two methods. They clearly prove the huge improvement that the triangulation method provides in the analytic calculations of multifold MB integrals.
%The MB representations of the massless conformal hexagon
%\begin{align}
%{I}_{\text{HX}}&=\frac{Q_6}{\Gamma(D/2-f)\Gamma(f)}
% \left( \prod_{i=1}^{9} \int_{-i\infty}^{+i\infty} w_i^{z_i} \frac{\Gamma(-z_i)}{2 \pi i} \right)
%\Gamma(D/2-f+z_1+...+z_9)\nonumber\\
%&\times\Gamma(b+z_1+z_5+z_8+z_9)\Gamma(c+z_2+z_6+z_7+z_8)
%\Gamma(d+z_3+...+z_6)\nonumber\\
%&\times\Gamma(e+f-D/2-z_4-...-z_9)\Gamma(-b-c-d-e+D/2-z_1-z_2-z_3-z_5-z_6-z_8)
%\label{HX}
%\end{align}
%and double-box
%\begin{align}
%{I}_{\text{DB}}&=\frac{Q_{3,3}\ u_8^{D/2-\ell}}{\Gamma(D/2-f)\Gamma(f)}
%\left( \prod_{i=1}^{9} \int_{-i\infty}^{+i\infty} w_i^{z_i} \frac{\Gamma(-z_i)}{2 \pi i} \right)
%\Gamma(D-f-\ell+z_1+...+z_9)\nonumber\\
%&\times\Gamma(b+z_1+z_5+z_8+z_9)\Gamma(c+z_2+z_6+z_7+z_8
%\Gamma(d+z_3+...+z_6)\nonumber\\
%&\times\Gamma(e+f+\ell-D-z_4-...-z_9)\Gamma(-b-c-d-e+D-\ell-z_1-z_2-z_3-z_5-z_6-z_8)\nonumber\\
%&\times\frac{\Gamma(\ell-D/2-z_7-z_8-z_9)}{\Gamma(-z_7-z_8-z_9)}
%\label{DB}
%\end{align}

One will also note that the hexagon and double-box were solved using the conic hulls method in \cite{Ananthanarayan:2020ncn} as sums of respectively $26$ and $44$ multivariable hypergeometric series, for generic values of the powers of their propagators satisfying the conformal constraint. However, due to computational limitations of the conic hulls approach only very few of all the possible series representations of these integrals could be derived. 
%Here, we apply the triangulation method to re-compute these MB integrals. To do so, we first find the point configuration of each integral. For Eq. \eqref{DB} it is a set of $16$ points of dimension $7$ which we write in the matrix-form as 
%\begin{align}
%A_\text{DB}=                                                     
%\left(
%\begin{array}{cccccccccccccccc}
% 1 & 1 & 1 & 1 & 1 & 1 & 1 & 1 & 1 & 1 & 0 & 0 & 0 & 0 & 0 & 0 \\
% 1 & 0 & 0 & 0 & 1 & 0 & 0 & 1 & 1 & 0 & 1 & 0 & 0 & 0 & 0 & 0 \\
% 0 & 1 & 0 & 0 & 0 & 1 & 1 & 1 & 0 & 0 & 0 & 1 & 0 & 0 & 0 & 0 \\
% 0 & 0 & 1 & 1 & 1 & 1 & 0 & 0 & 0 & 0 & 0 & 0 & 1 & 0 & 0 & 0 \\
% 0 & 0 & 0 & -1 & -1 & -1 & -1 & -1 & -1 & 0 & 0 & 0 & 0 & 1 & 0 & 0 \\
% -1 & -1 & -1 & 0 & -1 & -1 & 0 & -1 & 0 & 0 & 0 & 0 & 0 & 0 & 1 & 0 \\
% 0 & 0 & 0 & 0 & 0 & 0 & -1 & -1 & -1 & 0 & 0 & 0 & 0 & 0 & 0 & 1 \\
%\end{array}
%\right)\end{align}
%Similarly, one can find the point configuration of Eq.\eqref{HX} which is a set of $15$ points of dimension $6$.
Applying the triangulation method, we respectively obtain $194160$ and $243186$ series solutions for the hexagon and double box. Exploring the sets of these solutions using options such as \texttt{Cardinality} or \texttt{MaxCardinality} or \texttt{ShortestOnly}, we find simpler series solutions than those of \cite{Ananthanarayan:2020ncn}, as sums of $25$ hypergeometric series for both the hexagon and double box. These series solutions are presented in the ancillary Examples.nb \cite{MBConicHullsGit} notebook together with the resonant $D=4$ results.

\subsection{Higher-fold MB integrals: one-loop scalar massless $N$-point integral}
In order to show the huge improvement that the triangulation approach provides for calculations of MB integrals with a higher number of folds, we test it by considering the class of one-loop scalar massless $N$-point Feynman integrals with generic powers of the propagators, whose general MB representation for arbitrary $N$ is known for more than three decades (see Eq. (3.8) in \cite{Davydychev:1990jt} for the notation):
\begin{figure}
\centering
\includegraphics[scale=1.2]{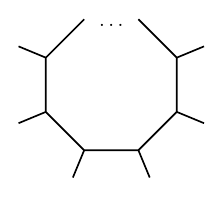}
\caption{One-loop $N$-point massless Feynman integral}
\end{figure}
\begin{align}
&J^{(N)}\left(\{\nu_j\}\vert\{p_j\};0\right)=\pi^{D/2}i^{1-D}\left(k_{1N}^2\right)^{D/2-\sum_i\nu_i}\frac{1}{\Gamma\left(D-\sum_i\nu_i\right)\prod_i\Gamma(\nu_i)}\frac{1}{(2\pi i)^{N(N-1)/2-1}}\nonumber\\
&\times\int\limits_{-i \infty}^{+i \infty} \cdots \int\limits_{-i \infty}^{+i \infty}\underset{(j,l)\neq(1,N)}{\prod_{j<l}}\left\{ds_{jl}\left(\frac{k_{jl}^2}{k_{1N}^2}\right)^{s_{jl}}\Gamma(-s_{jl})\right\}\Gamma\left(\sum_i\nu_i-D/2+\underset{(j,l)\neq(1,N)}{\sum_{j<l}}\sum_ls_{jl}\right) \nonumber \\
&\times \Gamma\left(D/2-\sum_i\nu_i+\nu_1-\underset{j\neq1}{\sum_{j<l}}\sum_{l}s_{jl}\right) \Gamma\left(D/2-\sum_i\nu_i+\nu_N-\underset{l\neq N}{\sum_{j<l}}\sum_{l}s_{jl}\right)
 \nonumber\\
& \times 
\prod_{i=2}^{N-1} \Gamma\left(\nu_i+\sum_{j<i}s_{ji}+\sum_{l>i}s_{il}\right) 
  \end{align}
The analytic expression of the $N=3$ case is well-known as a combination of four Appell $F_4$ double hypergeometric functions \cite{Boos:1990rg}. However, for $N\geq4$, due to the intricated structure of the poles in the MB integrand, it is indicated in \cite{Davydychev:1990jt} that it is considerably more complicated to obtain analytic results and, to the best of our knowledge, no such results have been published in the literature since then.

It is however easy to derive these results with our package for quite large values of $N$, as we have checked by considering the cases until $N=15$ (the latter one having a MB representation with 104 folds), see Table \ref{Speed2} for a few examples of the computation times. %and the number $M$ of series involved in the final expressions (we note that these results suggest that $M=$). 

\begin{table}[h]
\centering
    \begin{tabular}{ |p{1cm}|p{3cm}|p{3.5cm}|p{2.5cm}|}
\hline
%\multicolumn{5}{|c|c}{Calculation times comparison} \\
$N$ & Number of folds & Number of terms of the series solution & Computation time\\
\hline
4 & 5 & 11 & 0.384 sec.    \\
\hline
5  & 9 & 26 & 0.574 sec.    \\
\hline
10  & 44 & 1013 & 1.35 min. \\
\hline
13  & 77 & 8178 & 55.4 min. \\
\hline
15  & 104 & 32752  & 8.9 h.    \\
\hline
\end{tabular}
   \caption{Computation times of the one-loop $N$-point integral with the triangulation method. We show only the time taken to find a single triangulation and the corresponding set of poles associated to its series solution.}
   \label{Speed2}
\end{table}

Thanks to the master series of these series representations, which can also be obtained from our package, we have checked numerically these expressions (see \texttt{Examples.nb} notebook) against the direct numerical integration of the corresponding MB integrals using the \texttt{MB.m} package \cite{Czakon:2005rk}, for $N=4$ and $N=5$.

\section{Conclusion and Discussion\label{conclude}}

\bigskip
We have presented a new method for the analytic evaluation of multifold MB integrals, which is based on the triangulation of point configurations. 
Indeed, as explained in Section \ref{method}, to any given MB integrand one can associate a set of points, whose triangulations are in direct correspondence with series representations of the MB integral. After considering a simple illustrative example in Section \ref{example}, we have shown in Section \ref{Applications} how this approach considerably improves the speed of the computations compared to previous techniques, with the resulting fact that MB integrals with a very high number of folds can now be handled analytically in a reasonable calculation time. This is possible, due to the implementation of the triangulation technique in a new version of the \textit{Mathematica} package \texttt{MBConicHulls.wl} \cite{MBConicHullsGit} which was first developed in \cite{Ananthanarayan:2020fhl} (see also \cite{Banik:2022bmk}). We in addition showed how, by using appropriate options that we have added to the package, one can explore the space of series solutions, which can be huge for complicated MB integrals. In the particular cases of the hexagon and double-box conformal Feynman integrals, this allowed us to extract new and simpler series representations than the one previously obtained in \cite{Ananthanarayan:2020ncn}. %This also opens the way to important tests of the master series concept, as these two integrals, which are related by a differential equations, do not have the same total number of series representations. 
Other interesting studies of complicated integrals have been presented, as shown for instance with the case of the hard diagram of the two loop six edged Wilson loop in general kinematics and the one-loop $N$-point Feynman integrals for which new analytic results have been derived for the first time through this new computational technique.
\bigskip

\noindent {\bf Acknowledgements}

\medskip
We thank J. Rambau for useful discussions on TOPCOM. The work of S.B.~is supported by the Swiss National Science Foundation Grant Number PP00P21\_76884 . 
%\vspace{0.5cm}

\bigskip

\section{Appendix: Computer Implementation \label{package}}
We have implemented the triangulation method in an updated version of the \textit{Mathematica} package \texttt{MBConicHulls.wl} \cite{MBConicHullsGit} by introducing the new module \texttt{TriangulateMB[]} which we describe below. The other modules of the package mostly remain unchanged\footnote{The few minor changes can be found in the package using the  \texttt{?} command.} and their usage is explained in \cite{Ananthanarayan:2020fhl}. 
\mybox{
\textbf{\texttt{TriangulateMB[MBRepOut,Options[]]}}: This external module takes as input the output from the command \texttt{MBRep[]} of  \texttt{MBConicHulls.wl}  and first computes the point configuration associated to the MB integral. It then calls \texttt{TOPCOM} to find all the possible triangulations and prints the set of poles for each possible series solutions. 
\\\\
Below, we provide details about the input arguments and options for \texttt{TriangulateMB[]}.
\begin{itemize}
    \item \texttt{MBRepOut}: is the output of the \texttt{MBRep[]} function.
    \item Options:
    \begin{itemize}
    \item \texttt{MaxSolutions}: It specifies the maximum number of series solutions of the MB
integral that one wishes to evaluate. Its default value is \texttt{Infinity}.
    \item \texttt{MasterSeries}: It specifies whether to compute the master series for each of the series solutions found or not. Its default value is \texttt{True}.
    \item \texttt{TopComParallel}: It specifies whether to run \texttt{TOPCOM} in parallel or not. Its default value is \texttt{True}.
    \item \texttt{TopComPath}: It specifies the path to the \texttt{TOPCOM} executables. Its default value is \texttt{"/usr/local/bin/"}.
    \item \texttt{PrintSolutions}: It specifies whether to print the list of possible solutions along with their list of poles or not. Its default value is \texttt{True}.
    \item \texttt{ShortestOnly}: It specifies whether to only print the solution with shortest number of sets of poles or not. Its default value is \texttt{False}.
    \item \texttt{MaxCardinality}: It specifies the maximum length (\textit{i.e.} number of sets of poles) of the solutions to be considered. Its default value is \texttt{None}.
    \item \texttt{Cardinality}: It specifies the length of solutions which have to be considered. Its default value is \texttt{None}.
    \item \texttt{SolutionSummary}: It specifies whether to only print a summary of possible solutions along with their cardinality. Its default value is \texttt{False}.
    \item \texttt{QuickSolve}: It specifies whether to find only the quickest possible solution (only valid for non-resonant cases). This is useful for higher-fold MB integrals. Its default value is \texttt{False}.
    \end{itemize}
\end{itemize}
}
\bigskip

We next demonstrate the usage of \texttt{MBConicHulls.wl} by solving the $two$-fold MB representation of the Appell $F_1$ function considered in Sec. \ref{example}. As a first step, we load the package

\mybox{
\inbox{}{& 
{\tt SetDirectory[NotebookDirectory[]];} \\
&{\tt <<MBConicHulls.wl}};\\
\boxsplit\\
\outbox{2}{& {\tt Last Updated: 10$^{\text{th}}$ August, 2023}\\
& { {\tt \hspace{-0.35cm} Version 1.2 by S. Banik, S. Friot}}}\\}
assuming that the package is kept in the same directory as the notebook. We then input the MB representation in Eq. \eqref{F1MB} using \texttt{MBRep[]} as follows

\mybox{
\inbox{}{& 
{\tt MBRepOut= MBRep$\bigg[\dfrac{\text{Gamma[c]}}{\text{Gamma[a]}\text{Gamma[}\text{b}_\text{1}\text{]Gamma[}\text{b}_\text{2}\text{]}},\{\text{z}_1,\text{z}_2\},\{ \text{-u}_\text{1},\text{-u}_\text{2} \},$} \\
&{\tt $\{  \{ \text{-z}_\text{1}, \text{-z}_\text{2}, \text{a} + \text{z}_\text{1}+ \text{z}_\text{2}, \text{b}_\text{1} + \text{z}_\text{1} , \text{b}_\text{2} + \text{z}_\text{2} \} , \{ \text{c} + \text{z}_\text{1}+ \text{z}_\text{2} \} \} \bigg];$ }}\\
\boxsplit\\\\
\outbox{2}{& {\tt Non-Straight Contours.}\\
& { {\tt \hspace{-0.35cm} Time Taken 0.423482 seconds}}}\\
}
which, as we did not specify explicitly any values for the parameters $a, b_1, b_2$ and $c$, implicitly means that the MB representation has non-straight contours which separates the sets of poles of each gamma functions of the numerator of the integrand in the usual way. We then use \texttt{TriangulateMB[]} to give all possible regular triangulations and the list of series solutions.

\mybox{
\inbox{}{& 
{\tt TriangulateMBOut= TriangulateMB$ \big[ \text{MBRepOut,MaxSolutions} \to \text{3} \big] ;$} \\}\\
\boxsplit\\
\outbox{2}{& {\tt 
\tt The associated $\text{A-matrix}$ for this MB integral is $\; \begin{pmatrix}
1 & 1 & 1 & 0 & 0 \\
1 & 0 & 0 & 1 & 0 \\
0 & 1 & 0 & 0 & 1
\end{pmatrix}$
}\\
& { {\tt \hspace{-0.35cm} Degenerate case with 8 conic hulls}}}\\
\gibspace{}{& {\tt  Found 3 regular triangulations.}}\\
\gibspace{}{& {\tt The shortest series solution found is of cardinality 1.}}\\
\gibspace{}{& {\tt Cardinality 1:: Solution found 1. }}\\[-0.3cm]
\gibspace{}{& {\tt Cardinality 2:: Solution found 2.}}\\
\gibspace{}{& {\tt \textbf{Series Solution} 1:: Cardinality 1. Set of Poles ::$\{ \{ \text{n}_\text{1}, \text{n}_\text{2}  \} \}$ with }}\\[-0.3cm]
\gibspace{}{& {\tt master series characteristic list and variables $\{ \{ \text{n}_\text{1}, \text{n}_\text{2}  \} , \{ \text{-u}_\text{2},\text{-u}_\text{1} \} \}$.}}\\
\gibspace{}{& {\tt \textbf{Series Solution} 2:: Cardinality 2. Set of Poles ::$\{ \{ \text{n}_\text{1}, \text{-a} \text{-n}_\text{1} \text{-n}_\text{2}  \}$ }}\\[-0.3cm]
\gibspace{}{& {\tt $, \{ \text{n}_\text{1}, \text{-b}_\text{2} \text{-n}_\text{2}  \} \} $ with master series characteristic list and variables }}\\[-0.0cm]
\gibspace{}{& {\tt $\{ \{ \text{n}_\text{1}, \text{n}_\text{2}  \} , \{ \dfrac{1}{\text{-u}_\text{2}}  ,\text{-u}_\text{1} \} \}$.}}\\
\gibspace{}{& {\tt \textbf{Series Solution} 3:: Cardinality 2. Set of Poles ::$\{ \{ \text{-a} \text{-n}_\text{1} \text{-n}_\text{2} , \text{n}_\text{1}  \}$ }}\\[-0.3cm]
\gibspace{}{& {\tt $, \{ \text{-b}_\text{1} \text{-n}_\text{2} , \text{n}_\text{1} \} \} $ with master series characteristic list and variables }}\\[-0.0cm]
\gibspace{}{& {\tt $\{ \{ \text{n}_\text{1}, \text{n}_\text{2}  \} , \{ \dfrac{1}{\text{-u}_\text{1}}  ,\text{-u}_\text{2} \} \}$.}}\\
\gibspace{}{& {\tt Time Taken 0.491876 seconds}}
}

We used the option \texttt{MaxSolutions} to restrict the number of solutions to 3. The master series characteristic list and variables are also printed which is possible by exploiting the duality between the triangulation and conic hull approaches. Moreover, as $\Delta=0$ (see \cite{Ananthanarayan:2020fhl}), it prints that it is a degenerate MB representation (with $8$ conic hulls associated with it). In the next step we use \texttt{EvaluateSeries[]} in order to find the analytic expression of \textit{Series Solution 2} for some particular non-resonant values of the parameters (this is just for an illustration purpose because our package works for logarithmic resonant cases as well), as follows:

\mybox{
\inbox{}{& 
{\tt EvaluateSeriesOut= EvaluateSeries$ \big[ \text{TriangulateMBOut}, \{ \text{a} \to 1 , \text{b}_\text{1} \to \text{1/2}, $}\\
&{\tt $ \text{b}_\text{2} \to \text{1/3} , \text{c} \to \text{1/4}  \}, \, 2 \, \big] ;$ }}\\

\boxsplit\\
\outbox{2}{& { \tt The series solution is a sum of the following 2 series.
}\\
& { {\tt \textbf{Series Number} 1::   }}}\\
\gibspace{}{& {\tt $\dfrac{ (-1)^{\text{n}_\text{1}+\text{n}_\text{2}} \Gamma(\frac{1}{4}) \Gamma(\frac{1}{2}+\text{n}_\text{1})\Gamma(-\frac{2}{3}-\text{n}_\text{1}-\text{n}_\text{2}) \Gamma(1+\text{n}_\text{1}+\text{n}_\text{2}) (-u_1)^{\text{n}_\text{1}} (-u_2)^{-1-\text{n}_\text{1}-\text{n}_\text{2}}}{\sqrt{\pi}\Gamma(\frac{1}{3})\Gamma(-\frac{3}{4}-\text{n}_\text{2})\Gamma(1+\text{n}_\text{1})\Gamma(1+\text{n}_\text{2})}$}}\\
\gibspace{}{& {\tt valid for $\text{n}_\text{1} \geq \text{0} \, \, \text{\&\&} \, \, \text{n}_\text{2} \geq \text{0}$}}\\
\gibspace{}{& {\tt \textbf{Series Number} 2::  }}\\
\gibspace{}{& {\tt $ \dfrac{ (-1)^{\text{n}_\text{1}+\text{n}_\text{2}}\Gamma(\frac{1}{4}) \Gamma(\frac{1}{2}+\text{n}_\text{1})\Gamma(\frac{2}{3}+\text{n}_\text{1}-\text{n}_\text{2}) \Gamma(\frac{1}{3}+\text{n}_\text{2}) (-u_1)^{\text{n}_\text{1}} (-u_2)^{-\frac{1}{3}-\text{n}_\text{2}}}{\sqrt{\pi}\Gamma(\frac{1}{3})\Gamma(-\frac{1}{12}+\text{n}_\text{1}-\text{n}_\text{2})\Gamma(1+\text{n}_\text{1})\Gamma(1+\text{n}_\text{2})}$}}\\
\gibspace{}{& {\tt valid for $\text{n}_\text{1} \geq \text{0} \, \, \text{\&\&} \, \, \text{n}_\text{2} \geq \text{0}$}}\\
\gibspace{}{& {\tt Time Taken 0.640154 seconds}}
}
In the final step, we use \texttt{SumAllSeries[]} to perform a numerical evaluation of the series solution derived above for the chosen values $u_1 = -0.3$,  $u_2 = -10.1$ and for $n_1, n_2$ going from 0 to 15:
\mybox{
\inbox{}{& 
{\tt SumAllSeries$ \big[ \text{EvaluateSeriesOut}, \{ \text{u}_\text{1} \to \text{-0.3,}  \text{u}_\text{2} \to \text{-10.1} \}, \text{15} \big] $} }\\
\boxsplit\\
\outbox{2}{& {\tt Numerical Result :  -0.212049
}\\
& { {\tt \hspace{-0.3cm} Time Taken 0.244757 seconds }}}
}
and we cross-check the numerical value with the in-built \textit{Mathematica} function \texttt{AppellF1[]}.
\mybox{
\inbox{}{& 
{\tt AppellF1$ \bigg[ \text{1}, \dfrac{\text{1}}{\text{2}} , \dfrac{\text{1}}{\text{3}} , \dfrac{\text{1}}{\text{4}} , \text{-0.3}, \text{-10.1} \bigg] $} }\\
\boxsplit\\
\outbox{2}{& {\tt  -0.212049
}}
}


\begin{thebibliography}{99}


\bibitem{Smirnov:2012gma}
V.~A.~Smirnov,
\textit{``Analytic tools for Feynman integrals''},
Springer Tracts Mod. Phys. \textbf{250} (2012), 1-296
doi:10.1007/978-3-642-34886-0
%130 citations counted in INSPIRE as of 29 Jun 2020

\bibitem{Dubovyk:2022obc}
I.~Dubovyk, J.~Gluza and G.~Somogyi,
\textit{``Mellin-Barnes Integrals: A Primer on Particle Physics Applications,''}
Lect. Notes Phys. \textbf{1008} (2022), pp.
doi:10.1007/978-3-031-14272-7
[arXiv:2211.13733 [hep-ph]].

\bibitem{Gluza:2007rt}
J.~Gluza, K.~Kajda and T.~Riemann,
%``AMBRE: A Mathematica package for the construction of Mellin-Barnes representations for Feynman integrals,''
Comput. Phys. Commun. \textbf{177} (2007), 879-893
doi:10.1016/j.cpc.2007.07.001
[arXiv:0704.2423 [hep-ph]].

\bibitem{Ambre}
AMBRE webpage: http://prac.us.edu.pl/gluza/ambre

\bibitem{Belitsky:2022gba}
A.~V.~Belitsky, A.~V.~Smirnov and V.~A.~Smirnov,
%``MB tools reloaded,''
Nucl. Phys. B \textbf{986} (2023), 116067
doi:10.1016/j.nuclphysb.2022.116067
[arXiv:2211.00009 [hep-ph]].

\bibitem{Smirnov:1999gc}
V.~A.~Smirnov,
%``Analytical result for dimensionally regularized massless on shell double box,''
Phys. Lett. B \textbf{460} (1999), 397-404
doi:10.1016/S0370-2693(99)00777-7
[arXiv:hep-ph/9905323 [hep-ph]].

\bibitem{Tausk:1999vh}
J.~B.~Tausk,
%``Nonplanar massless two loop Feynman diagrams with four on-shell legs,''
Phys. Lett. B \textbf{469} (1999), 225-234
doi:10.1016/S0370-2693(99)01277-0
[arXiv:hep-ph/9909506 [hep-ph]].

\bibitem{Czakon:2005rk}
M.~Czakon,
%``Automatized analytic continuation of Mellin-Barnes integrals,''
Comput. Phys. Commun. \textbf{175} (2006), 559-571
doi:10.1016/j.cpc.2006.07.002
[arXiv:hep-ph/0511200 [hep-ph]].

\bibitem{Smirnov:2009up}
A.~V.~Smirnov and V.~A.~Smirnov,
%``On the Resolution of Singularities of Multiple Mellin-Barnes Integrals,''
Eur. Phys. J. C \textbf{62} (2009), 445-449
doi:10.1140/epjc/s10052-009-1039-6
[arXiv:0901.0386 [hep-ph]].

\bibitem{Kalmykov:2020cqz}
M.~Kalmykov, V.~Bytev, B.~A.~Kniehl, S.~O.~Moch, B.~F.~L.~Ward and S.~A.~Yost,
%``Hypergeometric Functions and Feynman Diagrams,''
doi:10.1007/978-3-030-80219-6\_9
[arXiv:2012.14492 [hep-th]].

%\cite{Vollinga:2004sn}
\bibitem{Vollinga:2004sn}
J.~Vollinga and S.~Weinzierl,
%``Numerical evaluation of multiple polylogarithms,''
Comput. Phys. Commun. \textbf{167}, 177 (2005)
doi:10.1016/j.cpc.2004.12.009
[arXiv:hep-ph/0410259 [hep-ph]].
%413 citations counted in INSPIRE as of 14 Aug 2023

\bibitem{Dubovyk:2019krd}
I.~Dubovyk, J.~Gluza and T.~Riemann,
%``Optimizing the Mellin-Barnes Approach to Numerical Multiloop Calculations,''
Acta Phys. Polon. B \textbf{50} (2019), 1993-2000
doi:10.5506/APhysPolB.50.1993
[arXiv:1912.11326 [hep-ph]].

\bibitem{Kalmykov:2016lxx}
M.~Y.~Kalmykov and B.~A.~Kniehl,
%``Counting the number of master integrals for sunrise diagrams via the Mellin-Barnes representation,''
JHEP \textbf{07} (2017), 031
doi:10.1007/JHEP07(2017)031
[arXiv:1612.06637 [hep-th]].

%\cite{Kalmykov:2012rr}
\bibitem{Kalmykov:2012rr}
M.~Y.~Kalmykov and B.~A.~Kniehl,
%``Mellin-Barnes representations of Feynman diagrams, linear systems of differential equations, and polynomial solutions,''
Phys. Lett. B \textbf{714}, 103-109 (2012)
doi:10.1016/j.physletb.2012.06.045
[arXiv:1205.1697 [hep-th]].
%36 citations counted in INSPIRE as of 02 Aug 2023


\bibitem{Srivastava}
H.~M.~Srivastava and P.~W.~Karlsson, \textit{``Multiple gaussian hypergeometric series''}, Ellis Horwood Series in Mathematics and Its Applications, 1985.  


\bibitem{Barnes}
E. W. Barnes,
%A new development of the theory of the hypergeometric functions
Proc. London Math. Soc. (2), 6 (1908), pp. 141-177.

\bibitem{W&W}
E. T. Whittaker and G. N. Watson,
 \textit{``A course of modern analysis''},
 Cambridge University press, 1902.

\bibitem{KdF}
  P.~Appell and J.~Kamp\'e de F\'eriet, 
  \textit{``Fonctions hyperg\'eom\'etriques et hypersph\'eriques - Polyn\^omes d'Hermite''},
  Gautiers-Villars et $\text{C}^{\text{ie}}$, 1926.
  
\bibitem{Erdelyi}
H. Bateman,  \textit{``Higher transcendental functions''}, Vol. 1, McGraw-Hill Book Company, New-York, Toronto, London, 1953.
  
\bibitem{Ananthanarayan:2020fhl}
B.~Ananthanarayan, S.~Banik, S.~Friot and S.~Ghosh,
%``Multiple Series Representations of N-fold Mellin-Barnes Integrals,''
Phys. Rev. Lett. \textbf{127} (2021) no.15, 151601
doi:10.1103/PhysRevLett.127.151601
[arXiv:2012.15108 [hep-th]].

\bibitem{Banik:2022bmk}
S.~Banik and S.~Friot,
%``Multiple Mellin-Barnes integrals with straight contours,''
Phys. Rev. D \textbf{107} (2023) no.1, 016007
doi:10.1103/PhysRevD.107.016007
[arXiv:2212.11839 [hep-ph]].

\bibitem{Ananthanarayan:2020ncn}
B.~Ananthanarayan, S.~Banik, S.~Friot and S.~Ghosh,
%``The Double Box and Hexagon Conformal Feynman Integrals,''
Phys. Rev. D \textbf{102} (2020) no.9, 091901
doi:10.1103/PhysRevD.102.091901
[arXiv:2007.08360 [hep-th]].
%%1 citations counted in INSPIRE as of 17 Nov 2020

\bibitem{Rambau:TOPCOM:2002}
J. Rambau, \textit{``TOPCOM: Triangulations of Point Configurations and Oriented Matroids,''}
 Mathematical Software - ICMS 2002 (Cohen, Arjeh M. and Gao, Xiao-Shan and Takayama, Nobuki, eds.), World Scientific (2002), pp. 330-340.

\bibitem{Davydychev:1990jt}
A.~I.~Davydychev,
%``Some exact results for N point massive Feynman integrals,''
J. Math. Phys. \textbf{32} (1991), 1052-1060
doi:10.1063/1.529383
%100 citations counted in INSPIRE as of 01 Sep 2020

\bibitem{Boos:1990rg}
E.~E.~Boos and A.~I.~Davydychev,
%``A Method of evaluating massive Feynman integrals,''
Theor. Math. Phys. \textbf{89} (1991), 1052-1063
doi:10.1007/BF01016805



\bibitem{MBConicHullsGit}
MBConicHulls webpage: https://github.com/SumitBanikGit/MBConicHulls


%\cite{DelDuca:2010zg}
\bibitem{DelDuca:2010zg}
V.~Del Duca, C.~Duhr and V.~A.~Smirnov,
%``The Two-Loop Hexagon Wilson Loop in N = 4 SYM,''
JHEP \textbf{05} (2010), 084
doi:10.1007/JHEP05(2010)084
[arXiv:1003.1702 [hep-th]].
%190 citations counted in INSPIRE as of 14 Aug 2023

%\cite{Ananthanarayan:2020xpd}
\bibitem{Ananthanarayan:2020xpd}
B.~Ananthanarayan, S.~Banik, S.~Friot and S.~Ghosh,
%``Massive One-loop Conformal Feynman Integrals and Quadratic Transformations of Multiple Hypergeometric Series,''
Phys. Rev. D \textbf{103}, no.9, 096008 (2021)
doi:10.1103/PhysRevD.103.096008
[arXiv:2012.15646 [hep-th]].
%11 citations counted in INSPIRE as of 02 Aug 2023

%\cite{Loebbert:2019vcj}
\bibitem{Loebbert:2019vcj}
F.~Loebbert, D.~M\"uller and H.~M\"unkler,
%``Yangian Bootstrap for Conformal Feynman Integrals,''
Phys. Rev. D \textbf{101}, no.6, 066006 (2020)
doi:10.1103/PhysRevD.101.066006
[arXiv:1912.05561 [hep-th]].
%39 citations counted in INSPIRE as of 14 Aug 2023

%\cite{Duhr:2022pch}
\bibitem{Duhr:2022pch}
C.~Duhr, A.~Klemm, F.~Loebbert, C.~Nega and F.~Porkert,
%``Yangian-Invariant Fishnet Integrals in Two Dimensions as Volumes of Calabi-Yau Varieties,''
Phys. Rev. Lett. \textbf{130}, no.4, 4 (2023)
doi:10.1103/PhysRevLett.130.041602
[arXiv:2209.05291 [hep-th]].
%17 citations counted in INSPIRE as of 08 Aug 2023




\end{thebibliography}
\end{document}